\documentclass[12pt]{iopart}
\usepackage{graphics,latexsym,epstopdf}
\usepackage{color}

\newcommand{\bra}[1]{\langle #1|}
\newcommand{\ket}[1]{|#1\rangle}

\begin{document}

\title{Distribution of entanglement in light-harvesting complexes  and their quantum efficiency }
\author{Francesca Fassioli}\ead{f.fassioli1@physics.ox.ac.uk}
\address{Department of Physics, University of Oxford, Clarendon Laboratory, Parks Road, Oxford OX1 3PU, UK}
\author{Alexandra Olaya-Castro}\ead{a.olaya@ucl.ac.uk}
\address{Department of Physics and Astronomy, University College London, Gower street, London WC1E 6BT, UK}

\begin{abstract}
Recent evidence of electronic coherence during energy transfer in photosynthetic antenna complexes has reinvigorated the discussion of whether coherence and/or entanglement has any practical functionality for these molecular systems. Here we investigate quantitative relationships between the quantum yield of a light-harvesting complex and the distribution of  entanglement among its components. Our study focusses on the {\it entanglement yield} or average entanglement surviving  a time scale comparable to the average excitation trapping time. As a prototype system we consider the Fenna-Matthews-Olson (FMO) protein of green sulphur bacteria  and show that there is an inverse relationship between the quantum efficiency and the average entanglement between distant donor sites. Our results suggest that long-lasting electronic coherence among distant donors might help modulation of the light-harvesting function. \end{abstract}

\maketitle
\section{Introduction}

The first step in photosynthesis is carried out by specialized pigment-protein complexes which absorb sunlight and, with near-unit efficiency, transfer the associated electronic excitation to reaction centres (RCs) where chemical energy storage begins. Numerous investigations have been devoted to elucidate the design principles of natural light-harvesting antenna \cite{review cogdgel, review schulten, simon09} and it is currently well known that a key element underlying their efficient functioning is the spatial arrangement of their pigment molecules together with their electronic interactions. However, the answer to whether electronic coherence in these systems could survive, first at physiological temperatures, and second, for long enough to influence the energy transfer process, had until recently, remained elusive. Remarkably, recent studies are providing evidence that, indeed, some of these fascinating molecular aggregates are designed to sustain quantum coherent transfer for longer than expected \cite{engel,lee07, gregnature10, engel10, gregjcpl10}, and that they can do so at temperatures as high as in biological conditions \cite{gregnature10,engel10}.  A direct implication suggested by these studies is that quantum coherence may be affecting light-harvesting processes in vivo \cite{rienk10}, but its exact role is yet to be understood.

The observation of coherent excited state dynamics in light-harvesting systems has therefore given a new impetus to the long-standing discussion regarding relationships between energy transfer efficiency and quantum features such as exciton delocalization (electronic coherence) and its coherent oscillatory behaviour \cite{makri08, mukamel97,gaab04}. In fact, several works have discussed highly efficient transport as an optimal interplay of coherent dynamics by which exciton phase relationships are preserved, and incoherent processes that destroy those relations  \cite{gaab04, alexandra08, masoud08, plenio08, rebentrost09, caruso09,cao,fassioli10,chin10}.  Within this framework, the role proposed for coherent transfer varies from being a strategy to overcome energy traps \cite{ishizaki09} to a mechanism that may speed up the search of the lowest energy level \cite{engel}.

The practical functionality of quantum coherent phenomena may, however, be subtler than efficiency maximization. For instance, the structure and arrangement of antenna complexes are not solely optimized for fast transfer but also incorporates the ability of modulating their function under different environmental conditions \cite{fleming,ruban,scheuring05,fassioli09,moulisova09,borrego99}. Then, one can ask how electronic coherence and its possible coherent evolution may help the requirement of having built-in mechanisms to modulate transfer efficiencies under different environmental parameters. Indeed, it may be that light-harvesting systems are able to take full advantage of unique quantum interference effects that include the possibility of efficiency regulation according to initial state conditions \cite{alexandra08, fassioli10} or through accumulative quantum phases in closed transfer pathways \cite{cao}.

An outstanding point of the recent experimental results is that the observed
electronic coherence expands several molecules across the whole antenna complex, including distant and weakly interacting pigments  \cite{ engel, gregnature10}.   This long-range electronic coherence has motivated recent studies to investigate electronic energy transfer from the perspective of quantum entanglement \cite{plenio,whaley,greg_np10}. An intriguing question is whether correlations or interferences between distant pigments play a part in the efficient functioning of a light-harvesting complex or whether they are just a consequence of exciton delocalization with no particular relation to transfer efficiencies.

In this work we use the framework of entanglement to quantify the strength of electronic coherences in a light harvesting system and investigate how such coherences are distributed among the molecular sites for different initial conditions and bath parameters.  Our study focuses on the effects of the coherences that survive a time scale that can be of biological relevance such as the average excitation trapping time.  These long-lasting electronic coherences are quantified by the {\it entanglement yield} defined as the average value of entanglement at times when trapping events are likely to occur.  As a prototype system we consider the FMO protein in green sulphur bacteria, where coherent dynamics was first observed at 77 K \cite{engel} and recently at 277 K  \cite{engel10}. Our results suggest that long-lived electronic coherence between well-separated pigments may modulate the efficiency profile as a function of the system-bath coupling. This modulation is manifested in an {\it inverse relationship} between quantum efficiency and the entanglement yield among distant pigments acting as excitation donors.  We discuss how the possible relation between efficiency and donor-donor coherence could be exploited in light-harvesting systems.

\section{Energy transfer dynamics in light-harvesting systems}
Under normal operating conditions energy transfer dynamics in light-harvesting complexes can be described under the assumption that there is at most a single excitation present \cite{fassioli09,sener05}. 
In this regime, the general form of the system's density matrix $\rho(t)$ takes the form:
\begin{equation}
\rho(t)=a_{00}(t)|0\rangle \langle 0|+\sum_{m=1}^N\sum_{n=1}^N a_{mn}(t)|m\rangle \langle n|\label{eq:density}
\end{equation}
where $|m\rangle=\ket{0..1_m....0}$ denotes the state with the excitation on site $m$ and all other chromophores (sites) in the ground state, $a_{00}(t)$ accounts for any losses that relax the system to the ground state $\ket{0}$, and $a_{mn}(t)$ describes the populations ($m=n$) and coherences ($m\neq n$) in the site basis.

The coherent evolution of $\rho(t)$ is dominated by the Hamiltonian describing the electronic states of $N$ interacting chromophores ($\hbar=1$):
\begin{equation}
H_S=~\sum_{m=1}^{N}\epsilon_m \sigma^{+}_{m} \sigma^{-}_{m}+\sum_{n>m}^N V_{mn}(\sigma^{+}_{m} \sigma^{-}_{n} + \sigma^{+}_{n} \sigma^{-}_{m})\label{Hs}
\end{equation}
where $\epsilon_m$ is the energy of the $m-th$ site, $V_{mn}$ describes the electronic coupling between chromophores $m$ and $n$, and $\sigma^{+(-)}_{m}$ is the usual creation (annihilation) operator of an excitation at site $m$.

The protein environment, which modulates site energy fluctuations and gives rise to decoherence, is usually modelled as a bath of harmonic oscillators linearly coupled to the electronic system \cite{fleming09review}. All the relevant information of the bath structure and system-bath interaction is then captured by the spectral density $J(\omega)$ and its associated site reorganization energy, $E_r=\int_0^{\infty} (J(\omega)/\omega) d\omega$. However the full  details of the coupling between the pigments and the surrounding protein environment are not fully characterized for these multichromophoric systems. In the FMO, for instance, there is an agreement that the reorganization energy values are in the moderate system-bath coupling regime with values between $E_r=35$  cm$^{-1}$  \cite{cho05,read08} 
and $E_r=25$ cm$^{-1}$ \cite{freiberg09}, but less consensus exists on the specific form of the bath spectral density and its characteristic correlation time. Reported bath correlation times,  span from 35 fs \cite{ishizaki09} to 166 fs \cite{cho05}, indicating that the energy transfer can happen under slow bath modulation where memory effects may be important. Importantly though, for bath-correlation times within this range, long-lived quantum coherence is expected at physiological conditions \cite{ishizaki09}.  The above discussion supports the idea that an understanding of the role of coherence and entanglement under weak system-bath coupling assumptions could give an insight into phenomena one can explore in more general conditions for which coherence survives. 

Here we consider energy transfer dynamics assuming generators of the Lindblad form $\dot{\rho}(t)=-i [H_S, \rho(t)] + L(\rho(t))$ which ensures complete positivity. First a quantum master equation, derived within the weak-coupling, Born-Markov and secular approximations \cite{fassioli10,breuerbook, masoud08} is considered:
\begin{equation}
L_1(\rho(t)) =  \sum_{\omega}\sum_{m,n}\gamma_{mn}(\omega) [A_n(\omega)\rho(t) A_m^\dag(\omega)-\frac{1}{2}\{A_m^{\dag}(\omega)A_n(\omega),\rho\}]\label{lindblad}
\end{equation}
This generator accounts for relaxation and dephasing in the eigenstates $\{\ket{\Psi}\}$ of $H_S$, with Lindbland operators given by $A_m(\omega)=\sum_{\varepsilon_{\Psi'}-\varepsilon_{\Psi}=\omega} c_m^*(\Psi)c_m(\Psi')|\Psi\rangle \langle \Psi'|$ and $\{\omega\}$  the frequency spectrum associated to  energy differences between single-excitation eigenstates. Note we have neglected the Lamb-shift type Hamiltonian. We assume that fluctuations at different sites can be correlated with  rates given by $\gamma_{mn}(\omega)=J_0(d_{mn}/\lambda_B)\gamma_{mm}(\omega)$  \cite{fassioli10} . Here $J_0$  is the Bessel function of first kind, $d_{mn}$ the distance between sites m and n, $\lambda_B$ the bath correlation length, and the single-site rate identical for all sites is $\gamma_{mm}(\omega)=\gamma(\omega)=2\pi {\mathcal J} (|\omega|)|N(-\omega)|$ with $N(\omega)$ being the thermal occupation number.  To compare our results with Ref.\cite{masoud08,alan}, we consider an ohmic spectral density of the form $J(\omega)=(E_r \omega/\omega_c) {\rm exp}(-\omega/\omega_c)$ with cutoff frequency $\omega_c$. The above model is limited to weak coupling and hence fails to reproduce the localization effects and the associated decay in the efficiency for large enough system-bath coupling \cite{ishizaki}.  

The second model  we consider is pure dephasing in the site basis \cite{rebentrost09, plenio08,gaab04}, which unlike (\ref{lindblad}), does not take into account the quantum nature of the bath. This model, however, describes qualitatively the transition from a purely coherent to incoherent regimes of energy transfer as function of the dephasing rate $\gamma_{k}$.  The generator  reads 
\begin{equation}
L_2(\rho(t)) =  \sum_{k}\gamma_{k} [A_k\rho(t) A_k^\dag-\frac{1}{2}\{A_k^{\dag}A_k,\rho\}]\label{puredephasing}
\end{equation}
with Lindblad operators in the site basis as $A_k=\ket{k}\bra{k}$. In our calculations, for simplicity, the dephasing rate is assumed identical for all sites, i.e.  $\gamma_{k}=\gamma_{deph}$.

The energy transfer dynamics governed by the coherent term $H_S$ and the bath-induced irreversible dynamics given by (\ref{lindblad}) or (\ref{puredephasing}), which preserve the number of excitations in the system, can be interrupted by dissipation to the ground state at any of the sites with a rate $\Gamma_m$ or excitation trapping at the RC with rate  $\kappa_{m}$.  Both of these processes are described by Lindblad generators of the form (\ref{puredephasing}) with operators $A_k=\sigma_k^{-}$ and rates $\gamma_{k}=\Gamma_k$ or $\gamma_{k}=\kappa_k$ for dissipation and trapping at the RC respectively. The effects of these generators are equivalent to the inclusion of a non-hermitian operator, $H_{\rm loss}=-(i/2)\sum_{m=1}^N(\Gamma_m +\kappa_{m})\, \sigma^{+}_{m}\sigma^{-}_{m}$, into the system Hamiltonian \cite{alexandra08,masoud08}.

\section{Quantum Coherence and Entanglement}
Quantum superpositions of the states of a composite system give rise to non-trivial  quantum features such as coherence and entanglement. These two phenomena are, however, not the same. While quantum coherence is generally associated to interference effects,
entanglement refers to non-classical correlations between distinguishable modes or subsystems of a multipartite complex \cite{horodecki09,vedral}. However, in scenarios
where the system's quantum evolution is restricted to a subspace with a single-excitation, quantum coherence can be not only a necessary but also a sufficient condition for the existence of entanglement \cite{whaley}.  Whether this entanglement can be used as an information resource is an ongoing discussion \cite{bartlett07}, 
though that is not the concern of this paper.  Our purpose 
here is to use entanglement as a framework to quantify electronic coherence and its distribution in a light-harvesting 
complex.

One of the most striking features of entanglement in the multipartite setting is the constrains in the way such quantum correlations can be distributed among subsystems \cite{coffman,osborne06,dawson05}. This property, known as monogamy, implies that entanglement cannot be freely shared among different objects, and therefore, the amount of quantum correlations that exist between two subsystems limits strongly the amount of entanglement they can share with a third party.  The monogamous nature of quantum correlations is expressed in strong inequalities bounding the way bipartite entanglement can be distributed, as originally demonstrated for pure states of three qubits by Coffman, Kundu, and Wootters \cite{coffman}. The generalization of such inequality  to the case of pure and mixed states of $N$ qubits has also been demonstrated \cite{osborne06}.  In particular, it has been shown that for a collection of $N$ qubits the distribution of bipartite entanglement, as measured by the squared concurrence or tangle \cite{wooters98}, satisfies the inequality:
\begin{eqnarray}
\sum_{m=1,m\neq n}^N\tau (\rho_{nm}) \leq \tau(\rho_{n:1\dots N})\equiv \tau(\rho_{n})\ . \label{eq:monogamy}
\end{eqnarray}
where $\tau(\rho_{n:1\dots N})$ refers to the tangle between the partition of site $n$ and remaining part of the system, and $\tau(\rho_{mn}(t))$ is the tangle of the reduced state $\rho_{mn}(t)$ of two sites $m$ and $n$.

In the context of the energy transfer dynamics we are interested in, the chromophores are considered as effective two-level systems or qubits and the two-qubit reduced dynamics $\rho_{mn}(t)$ is restricted to the zero- and single-excitation subspace. Hence, its associated concurrence is simply given by $C(\rho_{mn}(t))=2|a_{mn}(t)|$ and the corresponding tangle by $\tau(\rho_{mn}(t))=4|a_{mn}(t)|^2$, where $a_{mn}(t)$ are the off-diagonal elements of the density matrix in equation \ref{eq:density} representing electronic coherence, showing  that, in this case, entanglement is a measure of quantum coherence. The inequality in equation \ref{eq:monogamy}  consequently means that,  quantum coherence is also subject to certain shareability laws.

For our purpose of studying quantitative relationships between quantum yield and the spatial and temporal distribution of entanglement (or equivalently coherence) among the different chromophores in a light-harvesting complex, it  therefore appears important to consider a global measure that captures the monogamy relation.  In order to define such a measure one can notice that a summation over all tangles in equation (\ref{eq:monogamy}) leads to an inequality for the sum of all pairwise contributions i.e. $E_T(t)\leq \frac{1}{2} \sum_{n=1}^N\tau(\rho_{n})$ where
\begin{eqnarray}
E_T(t)=\sum_{m,n>m}\tau(\rho_{mn}(t))=4 \sum_{m,n>m}|a_{mn}(t)|^2\label{E_tot}
\end{eqnarray}
Notice that in the case of a pure state, $E_T(t)$ is closely related to the inverse participation ratio, often used as a measure of exciton delocalization \cite{mukamel97}. As it has been discussed before \cite{plenio,whaley}, the entanglement of $\rho(t)$ measured by $E_T(t)$ indicates that in the case where a single excitation is present in the system, the existence of non-vanishing off diagonal elements $a_{mn}(t)$ in equation \ref{eq:density} is sufficient condition for the presence of entanglement. However, we would like to highlight that here the use of this measure has been motivated by being one capturing the shareability of bipartite entanglement, important for our aim of relating efficiency to distribution of coherence.  As a first step in understanding such possible relations, we study the entanglement quantified by $E(t)$ and its associated yield $\phi$, which will be presented in the next section.

\section{Quantum yield and Entanglement yield}
The performance of a light-harvesting system can be assessed in terms of its efficiency, optimality and robustness \cite{sener05}. Here we focus on the quantum efficiency or quantum yield $\eta$, which measures the total probability that the excitation is trapped by the reaction centre as opposed to being radiatively dissipated at any of the network sites \cite{alexandra08}:
\begin{equation}
\eta =  \int_0^\infty \omega_{RC}(t)  dt
\end{equation}
where $\omega_{RC}(t)\equiv \sum_m \kappa_{m}\bra{m}\rho(t)\ket{m}$ is the probability density that the excitation is transferred from any of the sites to the RC.

In order to explore quantitative relationships between $\eta$ and the distribution of entanglement in space and time, here we concentrate on the amount of entanglement that lasts for the time scale of trapping events.  In particular, we focus on the expected value of entanglement at times when excitation trapping is likely. Formally, this corresponds to the average entanglement associated to the waiting time distribution defined by $\omega_{RC}(t)$. We define this quantity as the  {\it entanglement yield}:
\begin{equation}
\phi= \frac{1}{\eta}\int_0^\infty E(t) \omega_{RC}(t) dt\label{ent_yield}
\end{equation}
Different from previously used entanglement measures \cite{plenio,whaley}, the entanglement yield quantifies quantum correlations that may be impacting the transport properties on a time scale that compares with the average transfer time. In what follows we will investigate how this average entanglement gives insight into the modulating effect that long-lived electronic coherences may have on the quantum efficiency. \\

\begin{figure}
\centering
\resizebox{12.0cm}{!}{\includegraphics*{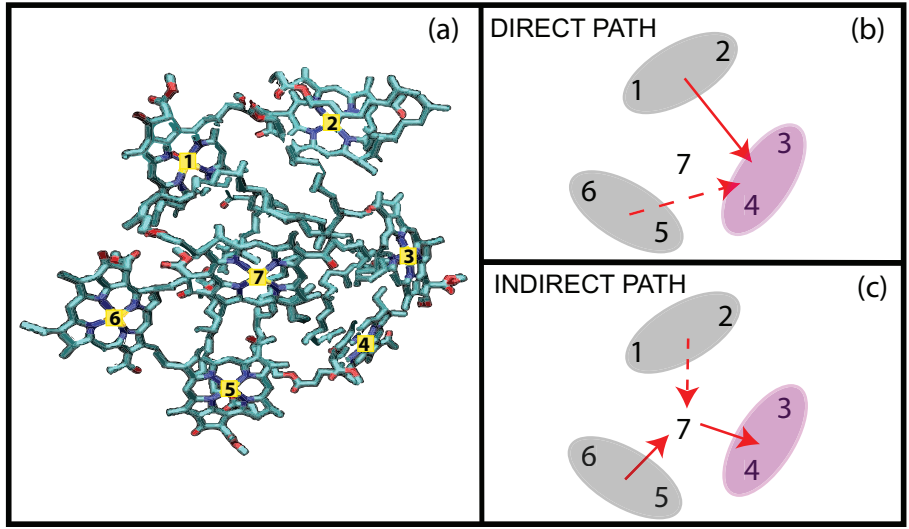}}
\caption{(a) Schematic representation of the monomeric subunit in the FMO complex. The position of each of the seven individual bacteriochlorophylls (BChls) is labeled from 1 to 7. (b-c) Most likely pathways of excitation transfer are shown as full arrows: from sites 1 and 2 the excitation is transferred primarily through a direct path (b) to the acceptor region, whereas from sites 5 and 6 it is most likely to be transferred through an indirect path (c) involving site 7. }
\label{fig1}
\end{figure}

\section{Entanglement yield distribution in FMO}
Contrary to what it could be assumed for antenna systems without symmetry, the spatial organization of pigments in a light-harvesting antenna is not random \cite{scheuring05,melkozernov}, and generally reflects the functionality associated with either individual pigments or a set of them. For instance, the FMO complex serves as a link between the chlorosome, which harvests the sunlight, and the RC. Theoretical \cite{adolphs06} and experimental \cite{wen09} studies support the view that each monomeric subunit of the FMO is oriented with sites 1 and 6 in closer contact with the baseplate chlorosome and with site 3 towards the RC. It is then likely that the excitation enters the FMO by this route, from where it is transferred to the target site 3. 

As a starting point to analyze the distribution of the entanglement yield (\ref{ent_yield}) in the FMO, we partition the system based on the predicted transfer pathways at 77 K \cite{cho05,brixner05}.  The results reported in \cite{cho05} indicate that there are two preferred non-cascade exciton relaxation pathways. One path involves mainly transfer from the strongly interacting pair 1 and 2 to the acceptor region defined by the sites 3 and 4. The second pathway takes the excitation from the sites 6 and 5 through site 7 until it reaches the acceptor region. We therefore define the pairs involving pigments 1-2 and 5-6 as donor dimers, whereas the 3-4 pair is denoted the acceptor dimer. The possible transfer pathways are classified between direct and indirect transfer pathways, depending on whether they exclude or include pigment 7 as illustrated in figure 1(b)-(c). With these partitions in mind, we investigate how the total entanglement $E_T(t)$ distributes along these pathways by partitioning $E_T(t)$ into donors, acceptors, intra-dimers, and mediator contributions:
\begin{eqnarray}
E_T(t)=E_{DD}+E_{AD}+E_{dim}+E_{7,rest}
\end{eqnarray}
where
\begin{eqnarray}
E_{DD} & \equiv & \tau_{1,5}+\tau_{1,6}+\tau_{2,5}+\tau_{2,6}\label{E_DD} \\
E_{AD} & \equiv & \sum_{i=1,2,5,6} \tau_{i,3}+\tau_{i,4}\\ 
E_{dim} & \equiv & \tau_{1,2}+\tau_{3,4}+\tau_{5,6}\\ 
E_{7,rest} & \equiv & \sum_{i\neq 7} \tau_{7,i}\label{E_7} 
\end{eqnarray}
Notice that the inter-dimer entanglements donor-donor $E_{DD}$ and acceptor-donor $E_{AD}$, account primarily for long-range coherences while the mediator entanglement $E_{7,rest}$ and the intra-dimer entanglement $E_{dim}$ quantify mainly short-range coherences. Their corresponding entanglement yields are defined as $\phi_{DD}$, $\phi_{AD}$,  $\phi_{dim}$ and $\phi_{7,rest}$ and are calculated through (\ref{ent_yield}). Note that for ease of notation $\tau_{m,n}\equiv\tau(\rho_{mn})$ in (\ref{E_DD}-\ref{E_7}).\\

In (\ref{lindblad}) we assume the same parameters as in \cite{masoud08} with a thermal bath at room temperature $T=293 K$ and a cutoff frequency $\omega_c=150$ cm$^{-1}$. The excitation is assumed to be transferred to the RC from site 3 with a rate $\kappa_3=1$ ps$^{-1}$ and radiative relaxation occurs with a rate of $\Gamma_{m}\equiv\Gamma=1$ ns$^{-1}$, identical for all pigments \cite{sener05}. The site energies and electronic couplings in (\ref{Hs}) have been taken from table 1 in reference \cite{cho05}.

\subsection{Entanglement survival}
\begin{figure}
\centering
\resizebox{15.0cm}{!}{\includegraphics*{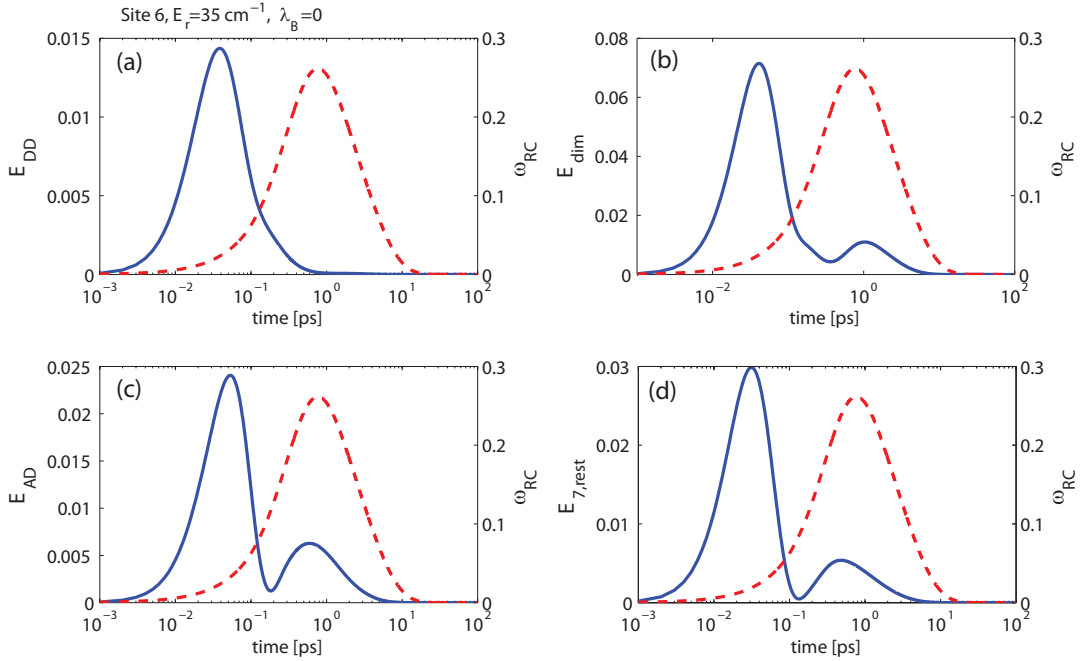}}
\caption{{\it  Weak-coupling and secular approximation model}. Time evolution of the contributions to the total entanglement (--) and RC trapping probability density (- -) for excitation initially localized on BChl 6,  at $T=293$ K, $E_r=35$ cm$^{-1}$ and $\lambda_B=0$. (a) $E_{DD}$, distant-donors contribution  (b) $E_{dim}$, intra-dimer contribution (c) $E_{AD}$, acceptor-donor contribution and (d) $E_{7,rest}$, site 7 contribution.}
\label{fig_time}
\end{figure}

In order to illustrate the entanglement dynamics,  we plot in figure \ref{fig_time} the time evolution of the different entanglement  contributions i.e. $E_{DD}$, $E_{AD}$, $E_{dim}$ and $E_{7,rest}$, assuming the excitation initially localized on site 6 and decoherence described by model (\ref{lindblad}). Here we consider $E_r={\rm 35\, cm^{-1}}$ and uncorrelated energy fluctuations ($\lambda_B=0$). For comparison, the  trapping probability density $\omega_{RC}(t)$ (dashed line) is also shown.  Several interesting features can be drawn from figure \ref{fig_time}. Firstly, the total entanglement, which is the sum of these four contributions, is predicted to have a maximum value first at around $0.04$~ps and then at around {$0.8$}~ps, close to where the trapping probability reaches a maximum too. These time scales of non-vanishing entanglement compare well with the experimental observations of coherence surviving at least $0.3$~ps at 277 K \cite{engel10}.  Secondly, the entanglement values that we are interested in correspond to the overlap with the waiting time distribution $\omega_{RC}(t)$, that is, the entanglement at times where there is non-vanishing trapping probabilities, i.e. between $t=0.01$~ps and $t=2$~ps  (see figure \ref{fig_time}).  Notice that although the entanglement values are small, there is a considerable overlap in time where both  $\omega_{RC}(t)$ and the entanglement contributions have values different from zero.  Interestingly, except for $E_{DD}$,  the entanglements $E_{AD}$, $E_{7,rest}$ and $E_{dim}$ peak closely where $\omega_{RC}(t)$ does. With the exception of the second peak in $E_{dim}$,  these results are similar to the case where the excitation is initially localized on site 1 (results not shown).  Given that the quantum yield is near unity with the parameters considered (figure \ref{fig3}(a)), we can conjecture that larger efficiencies are associated with both lower yet not zero values of $E_{DD}$ and larger values of $E_{AD}$ and $E_{7,rest}$ at typical trapping times.  In order to explore this hypothesis, we now turn to discuss the entanglement yield. 

\subsection{Efficiency and donor-donor entanglement yield }
\begin{figure}
\centering
\resizebox{14.0cm}{!}{\includegraphics*{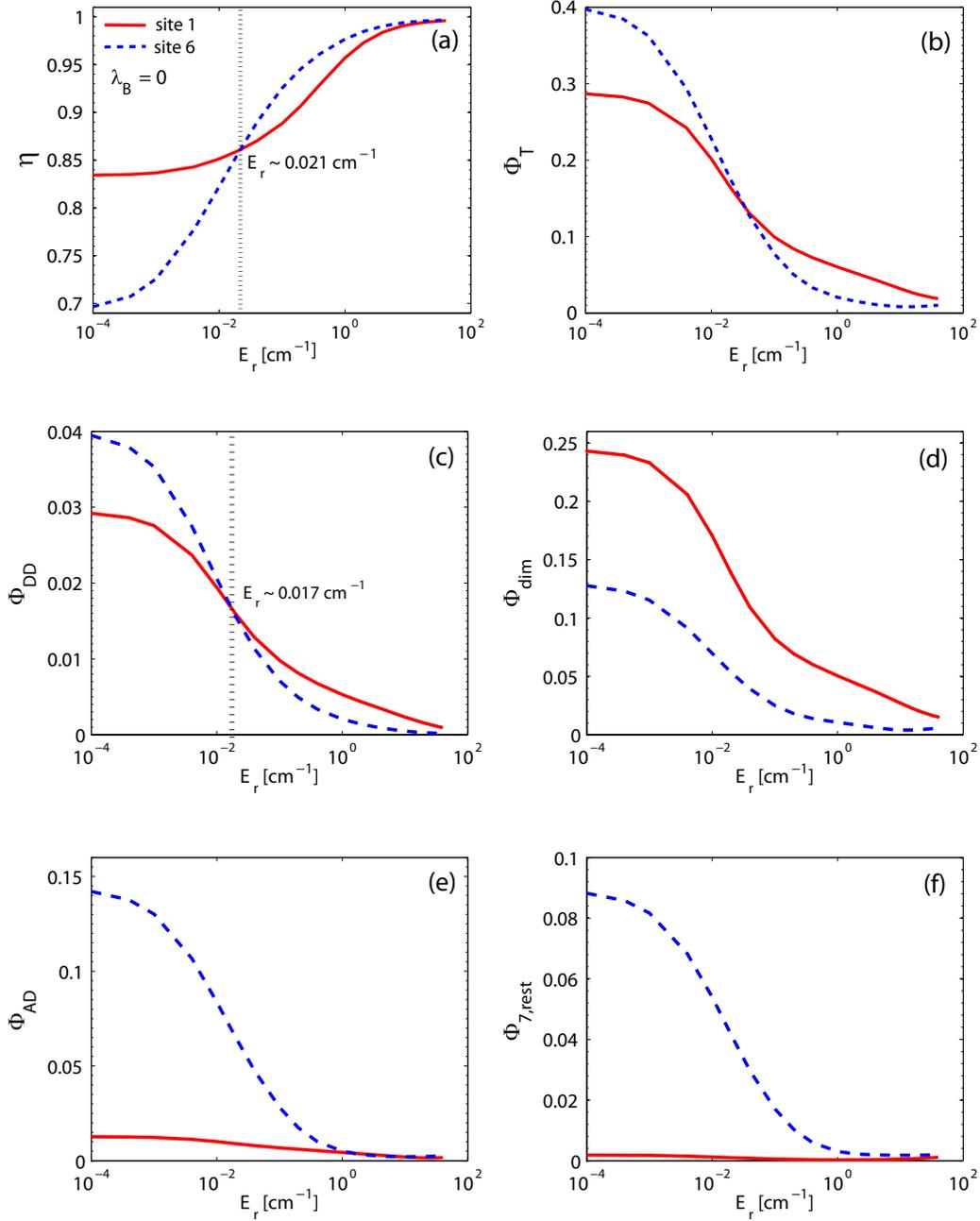}}
\caption{{\it  Weak-coupling and secular approximation model}. (a) $\eta$, quantum yield (b) $\phi_T$, total entanglement yield (c) $\phi_{DD}$, donor-donor entanglement yield (d) $\phi_{dim}$, intra-dimer entanglement yield (e) $\phi_{AD}$, acceptor-donor entanglement yield and (f) $\phi_{7,rest}$, site 7 entanglement yield, as a function of reorganization energy, for the initial states localized on site 1 (--) and site 6 (- -), for $T=293$ K and $\lambda_B=0$. }
\label{fig3}
\end{figure}

Quantitative relations between efficiency and the different contributions to the total entanglement yield were investigated by comparing their values under different bath parameters and various initial conditions. The initial states considered are excitation localized on either site 1 or  6. 
Under decoherence given by (\ref{lindblad}) with uncorrelated fluctuations ($\lambda_B=0$),  these two initial states exhibit, in general, different efficiencies as a function of the system-bath coupling quantified by $E_r$ (see figure \ref{fig3}(a)). As previously reported \cite{masoud08}, this model predicts a monotonic increase of $\eta$ with increasing $E_r$, which is a behaviour  valid only for weak system-bath couplings \cite{ishizaki}. Therefore, the maximum $E_r$ we consider is $E_r={\rm 50\, cm^{-1}}$. As interaction with the environment is stronger, coherence is destroyed and, consequently, the total entanglement yield decreases monotonically as shown in figure \ref{fig3}(b). Interestingly, the interplay between coherent and incoherent processes prompts a crossing of the efficiency profiles for the two different states at around $E_r\approx 0.021$ cm$^{-1}$, indicating that one  state is more efficient than the other one depending on the coupling to the environment. A similar crossing is observed for the total entanglement yield but at a slightly larger value i.e. $E_r\approx 0.035$ cm$^{-1}$. 
However, such crossing is not necessarily exhibited by all the contributions to the total entanglement yield as it can be seen in figure \ref{fig3}(c)-(f). In particular, notice that  $\phi_{DD}$ shown in figure \ref{fig3}(c) is the only contribution featuring a profile crossing at around $E_r\approx 0.017$ cm$^{-1}$. The most important feature of figure \ref{fig3} is the fact that for a large range of $E_r$ parameters an inverse relationship between  $\eta$ and $\phi_{DD}$ is suggested. For instance, if we compare the efficiency  and entanglement yield values for the two initial states at  $E_r={\rm 10\, cm^{-1}}$, we can see that they satisfy $\eta^1(E_r={\rm 10\, cm^{-1}}) < \eta^6(E_r={\rm 10\, cm^{-1}})$  while $\phi_{DD}^1(E_r={\rm 10\, cm^{-1}}) > \phi_{DD}^6\,(E_r={\rm 10\, cm^{-1}})$. This supports the idea that higher efficiencies are associated to lower values in donor-donor entanglement.

Importantly, the trends calculated for $\phi_{dim}$, $\phi_{AD}$ and $\phi_{7,rest}$ as function of bath reorganization energy, $E_r$ (see figure \ref{fig3}(d)-(f)) illustrate the differences in the distribution of entanglement among sites depending on the initial state. For example, the average entanglement of mediator site 7 and the rest, $\phi_{7,rest}$, is always larger for the case when the excitation is initially localized on site 6 than for the initial state at site 1 since, as discussed above, energy migration from site 6 is largely mediated by site 7 (see figure 1(c)).  Similarly, the donor-acceptor entanglement yield, $\phi_{AD}$ is in general larger for initial state at site 6 than 1 as interactions between sites 5 and 4 are larger than between sites 1 or 2 and sites 3 or 4 (cf. hamiltonian given in \cite{cho05}).  In contrast, $\phi_{dim}$  is always greater for the initial state at site 1 consistent with the fact that energy transfer from this site has a large contribution of the coherence associated to the strong interacting 1-2 pair. We would like to point out that so far we have not found clear relationships between efficiency and the contributions $\phi_{dim}$, $\phi_{AD}$ and $\phi_{7,rest}$.  

\begin{figure}
\centering
\resizebox{14.0cm}{!}{\includegraphics*{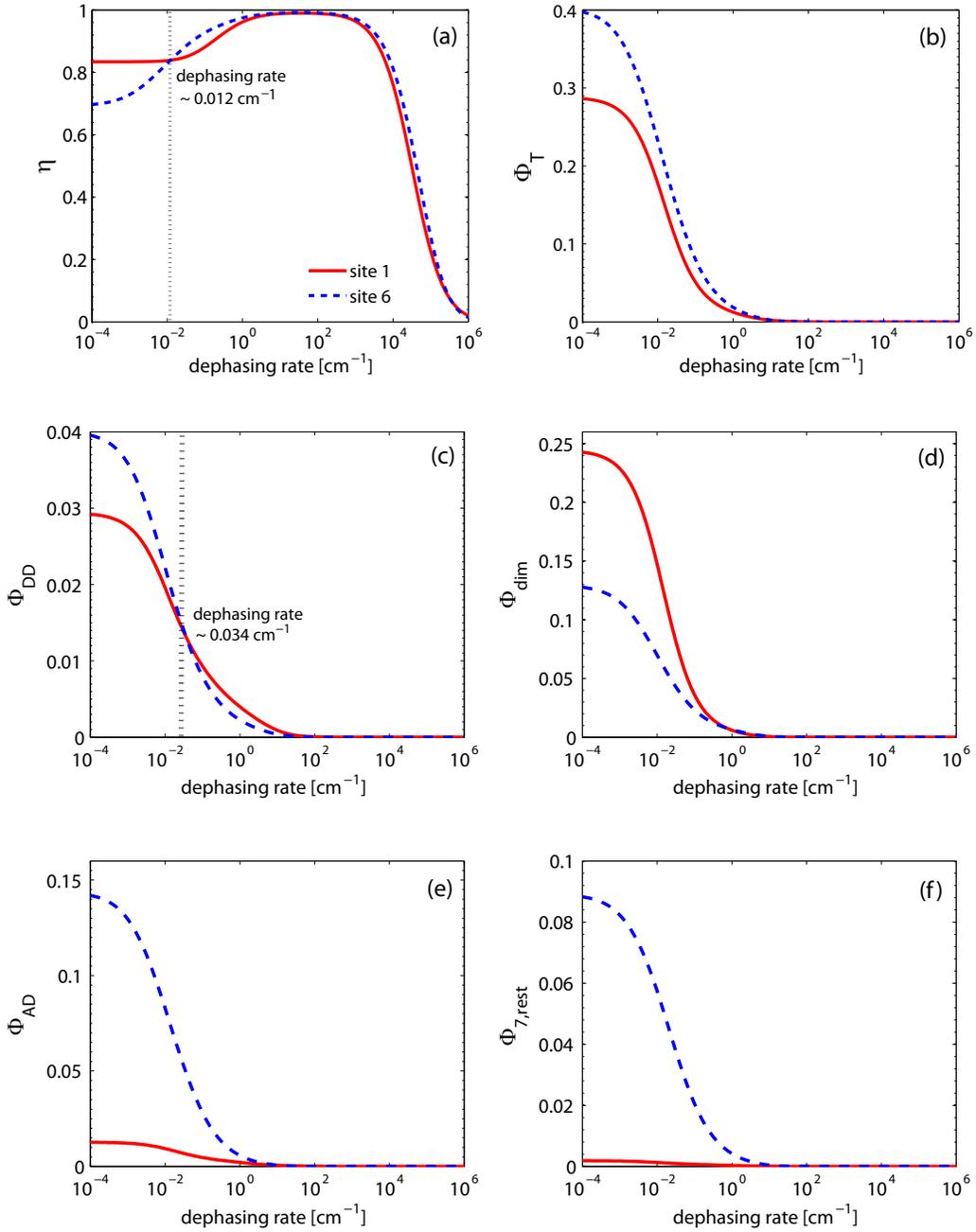}}
\caption{{\it Pure dephasing model.} (a) $\eta$, quantum yield (b) $\phi_T$, total entanglement yield  (c) $\phi_{DD}$, donor-donor entanglement yield (d) $\phi_{dim}$, intra-dimer entanglement yield (e) $\phi_{AD}$, acceptor-donor entanglement yield and (f) $\phi_{7,rest}$, site 7 entanglement yield, as a function of dephasing rate for the initial states localized on site 1 (--) and site 6 (- -). }
\label{fig5}
\end{figure}

The majority of the above analyses and, in particular, the identified inverse relationship between $\phi_{DD}$ and $\eta$ hold when decoherence is given by local random fluctuations described by the pure dephasing model (\ref{puredephasing}) as illustrated in figure \ref{fig5}. The quantum yield and the total entanglement yield as functions of the dephasing rate $\gamma_{deph}$  are shown in figures \ref{fig5}(a) and \ref{fig5}(b) respectively, while the different contributions to the entanglement yield are depicted in figures \ref{fig5}(c)-(f). As reported in \cite{rebentrost09, plenio08}, the quantum yield has a non-monotonic behaviour as the dephasing rate increases (see  figure \ref{fig5}(a)), being qualitatively consistent with a transition from coherent to incoherent transfer.  The efficiency profiles for the two initial states cross (figure \ref{fig5}(a)), similarly to the weak-coupling results (figure \ref{fig3}(a)). Nevertheless, there is no intersection in their total entanglement yields (compare figures \ref{fig3}(b) and \ref{fig5}(b)).  Entanglement yield crossing is however observed in the donor-donor contribution (compare figures \ref{fig3}(c) and \ref{fig5}(c)), pointing out again the inverse relationship between quantum efficiency and the average of donor-donor entanglement at trapping times. 

The main difference between the two decoherence models here considered is the crossing of the profiles of the total entanglement yield for the two initial states, which for the secular case is observed where coherence has not vanished (see figure \ref{fig3}(a)), while in the pure dephasing model there is no such crossing before coherence is practically zero (see figure \ref{fig5}(c)). We believe these differences arise from the bath features that are considered by each model. ÊIt has been discussed that the phonon-assisted transport properties in FMO complex are greatly dominated by the temperature-independent spontaneous emission of energy into the phonon bath \cite{masoud08}. These fluctuations are accounted for within the secular approximation while they are completely neglected by the classical bath approximation in pure-dephasing models. Despite of the differences, however, the suggested inverse relationship between the efficiency and the average of donor-donor entanglement $\phi_{DD}$ holds for the two decoherence models. 

\begin{figure}
\centering
\resizebox{14.0cm}{!}{\includegraphics*{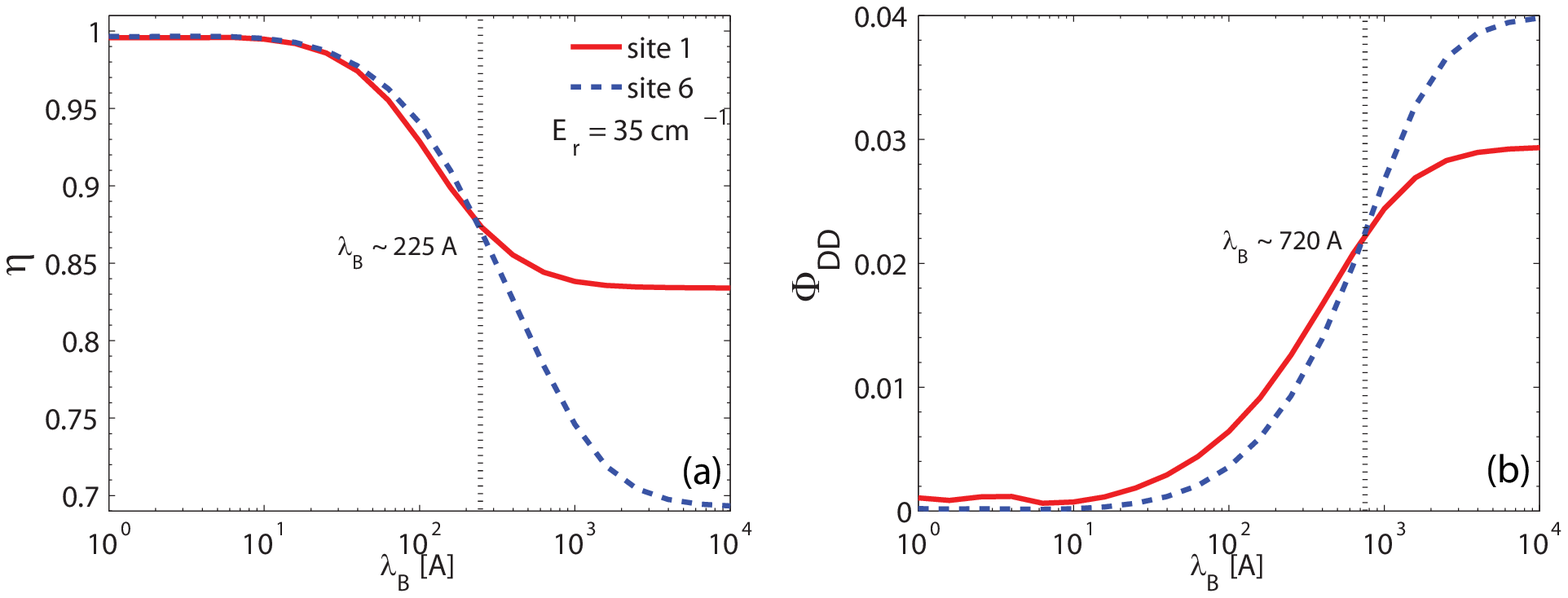}}
\caption{{\it  Weak-coupling and secular approximation model, including spatial correlations}. (a) Quantum yield and (b) $\phi_{DD}$, donor-donor entanglement yield as a function of bath correlation length, for the initial states localized on site 1 (--) and site 6 (- -), for $T=293$ K and $E_r=35$ cm$^{-1}$.}
\label{fig7}
\end{figure}

\subsection{Entanglement yield under spatially correlated fluctuations} 
Long-range spatial correlations have been suggested as an important mechanism helping quantum coherence to survive at high-temperatures in light-harvesting systems \cite{engel, gregnature10, engel10}. This  hypothesis is supported by recent studies showing that for a single donor-acceptor pair environmental fluctuations that are correlated in space lead to a high critical temperature for which a transition from coherent to incoherent transfer occurs \cite{nazir09}.
In this section we explore the robustness of the above results when bath-induced energy fluctuations of different pigments are correlated \cite{fassioli10}. In particular we are interested in determining whether the suggested inverse relationship between $\eta$ and $\phi_{DD}$ still holds. We therefore fix the reorganization energy to $E_r={\rm 35\, cm^{-1}}$ and compare the profiles of quantum and entanglement yields as functions of the degree of bath correlation by varying $\lambda_B$ in (\ref{lindblad}). 
In the presence of spatial correlations the efficiency for the initial states considered decreases \cite{alan} and distinguishability of the initial state becomes evident \cite{fassioli10} (see figure \ref{fig7}(a)). As correlations are included the coherent nature of the transfer dominates and the entanglement yield increases. Importantly, the inverse relationship between efficiency and the average entanglement among donors is again observed (see figure \ref{fig7}(b)).

\section{Concluding Remarks}
A key challenge in the theory of photosynthetic energy transfer is formulating possible quantitative relations between electronic coherence and the light-harvesting function. In this work we have approached this challenge by investigating relationships between the transfer efficiency and the entanglement present in the system at times when trapping events are likely to occur. We have shown that in the FMO complex and under weak system-bath coupling assumptions, this long-lived coherence is spatially distributed in such a way that the average entanglement among donor dimers exhibits an inverse relationship with the quantum yield.   This feature is observed under different decoherence models and a wide range of bath parameters including room temperature, suggesting that donor-donor correlations may be exploited for efficiency modulation.  Further investigation is however needed in order to asses the validity of this quantitative  relation in more general theories of energy transfer that are not limited to the weak-coupling limit here explored  \cite{ishizaki,nazir09,jang08,jang09}.

The ability to regulate transport properties under different environmental conditions is crucial for robust photosynthetic systems.  Under high light-conditions, for instance, photosynthetic systems switch on photo-protection mechanisms that allow them to control (reduce) how fast excitation energy is converted into chemical energy \cite{fleming,ruban}. The mechanism here described by which a convenient distribution of electronic coherence among pigments could allow control of the efficiency profile, may prove useful in adjusting transport properties under variations of light-intensity.  It will therefore be interesting to investigate correlations between efficiency and the spatio-temporal distribution of coherence when pigment-protein complexes are exposed to different light intensities.

\ack
We are grateful to Avinash Kolli, Dara McCutcheon and Ahsan Nazir for fruitful discussions. We would also like to thank  Dan Browne and Gregory D Scholes for helpful suggestions and critical reading of this manuscript. FF thanks CONICYT (Chile) and EPSRC for funding. AOC gratefully acknowledges EPSRC grant EP/G005222/1 for financial support.


\section*{References}
\begin{thebibliography}{10}
\bibitem{review cogdgel} Cogdell R J, Gall A and K\"ohler J 2006 {\it Q. Rev. Biophys.} {\bf 39} 227
\bibitem{review schulten} Hu X, Ritz T, Damjanovic A, Autenrieth F and Schulten K 2002 {\it Q. Rev. Biophys.} {\bf 35} 1
\bibitem{simon09} Scheuring S and Sturgis J N 2009 {\it Photosynth. Res.} {\bf 102} 197
\bibitem{engel} Engel G S, Calhoun T R, Read E L, Ahn T-K, Man\v{c}al T, Cheng Y-C, Blankenship R E and Fleming G R 2007 {\it Nature} {\bf 446} 782 
\bibitem{lee07} Lee H, Cheng Y-C and Fleming G R 2007 {\it Science} {\bf 316} 1462
\bibitem{gregnature10} Collini E, Wong C Y, Wilk K E, Curmi P M G,  Brumer P and  Scholes G D 2010 {\it Nature} {\bf 463} 644
\bibitem{engel10} Panitchayangkoon G, Hayes D, Fransted K A, Caram J R, Harel E, Wen J, Blankenship R E and Engel G S 2010 arXiv:1001.5108v1
\bibitem{gregjcpl10} Scholes G D 2010 {\it J. Phys. Chem. Lett.} {\bf 1} 2
\bibitem{rienk10} van Grondelle R and Novoderezhkin V I 2010  {\it Nature} {\bf 463} 614 
\bibitem{makri08} Ray J and Makri N 1999 {\it J. Phys. Chem. A} {\bf 103} 9417
\bibitem{mukamel97} Meier T, Chernyak V and Mukamel S 1997 {\it J. Phys. Chem. B} {\bf 101} 7332
\bibitem{gaab04} Gaab K M and Bardeen C J 2004 {\it J. Chem. Phys.} {\bf 121} 7813 
\bibitem{alexandra08} Olaya-Castro A, Lee C F, Fassioli Olsen F and Johnson N F 2008 {\it Phys. Rev. B} {\bf 78} 085115
\bibitem{masoud08} Mohseni M, Rebentrost P, Lloyd S and Aspuru-Guzik A 2008 {\it J. Chem. Phys.} {\bf 129} 174106
\bibitem{plenio08} Plenio M B and Huelga S F 2008 {\it New J. Phys.} {\bf 10} 113019 
\bibitem{rebentrost09} Rebentrost P, Mohseni M, Kassal I, Lloyd S and Aspuru-Guzik A 2009 {\it New J. Phys.} {\bf 11}  033003
\bibitem{caruso09} Caruso F, Chin A W, Datta A, Huelga S F and Plenio M B 2009 {\it J. Chem.  Phys} {\bf 131} 105106
\bibitem{cao} Cao J and Silbey R J 2009 {\it J. Phys. Chem. A} {\bf 113} 13825
\bibitem{fassioli10} Fassioli F, Nazir A and Olaya-Castro A 2010 {\it J. Phys. Chem. Lett.} {\bf 1} 2139
\bibitem{chin10} Chin A W, Datta A, Caruso F, Huelga S F and Plenio M B 2010 {\it New J. Phys.} {\bf 12} 065002
\bibitem{ishizaki09} Ishizaki A and Fleming G R 2009 {\it Proc. Natl. Acad. Sci. USA} {\bf 106} 17255
\bibitem{fleming}  Ahn T K,  Avenson T J,  Ballottari M, Cheng Y-C, Niyogi K K, Bassi R and Fleming G R 2008 {\it Science} {\bf 320} 794
\bibitem{ruban} Ruban A V, Berera R, Ilioaia C, van Stokkum I H M, Kennis J T M, Pascal A A, van Amerongen H, Robert B, Horton P and van Grondelle R 2007 {\it Nature} {\bf 450} 575
\bibitem{scheuring05} Scheuring S and Sturgis J N 2005 {\it Science} 2005 {\bf 309} 484
\bibitem{fassioli09} Fassioli F, Olaya-Castro A, Scheuring S, Sturgis J N and Johnson N F 2009 {\it Biophys. J.} {\bf 97} 2464
\bibitem{moulisova09} Moulisov\`a V, Luer L, Hoseinkhani S, Brotosudarmo T H P, Collins A M, Lanzani G, Blankenship R E and Cogdell R J 2009 {\it Biophys. J.} {\bf 97} 3019
\bibitem{borrego99} Borrego C M, Gerola P D, Miller M and Cox R P (1999) {\it Photosynth. Res.} {\bf 59} 159
\bibitem{plenio} Caruso F, Chin A W, Datta A, Huelga S F and Plenio M B 2009 arXiv:0912.0122v1
\bibitem{whaley} Sarovar M, Ishizaki A, Fleming G R and Whaley K B 2010 {\it Nature Physics} {\bf 6} 462
\bibitem{greg_np10} Scholes G D 2010 {\it Nature Physics} {\bf 6} 402
\bibitem{sener05} Sener M and Schulten K 2005 {\it Energy Harvesting Materials}, edited by David L. Andrews (Singapore: World Scientific)
\bibitem{fleming09review} Cheng Y-C and Fleming G R 2009 {\it Annu. Rev. Phys. Chem.} {\bf 60} 241
\bibitem{cho05} Cho M, Vaswani H M, Brixner T, Stenger J and Fleming G R 2005 {\it J. Phys. Chem. B} {\bf 109} 10542
\bibitem{read08} Read E L, Schlau-Cohen G S, Engel G S, Wen J, Blankenship R E and Fleming G R 2008 {\it Biophys. J.} {\bf 95} 847 
\bibitem{freiberg09} Freiberg A, R\"atsep M, Timpmann K and Trinkunas G 2009 {\it Chem. Phys.} {\bf 357} 102
\bibitem{breuerbook} Breuer H -P and Petruccione F 2002 {\it The Theory of Open Quantum Systems} (New York: Oxford University Press)
\bibitem{alan} Rebentrost P, Mohseni M and Aspuru-Guzik A 2009 {\it J. Phys. Chem. B} {\bf 113} 9942 
\bibitem{ishizaki} Ishizaki A and Fleming G R 2009 {\it J. Chem. Phys.} {\bf 130} 234110
\bibitem{horodecki09} Horodecki R, Horodecki P, Horodecki M and Horodecki K 2009 {\it Rev. Mod. Phys.} {\bf 81} 865
\bibitem{vedral} Vedral V 2008 {\it Nature} {\bf 453} 1004
\bibitem{bartlett07} Bartlett S D, Rudolph T and Spekkens R W 2007 {\it Rev. Mod. Phys.} {\bf 79} 555
\bibitem{coffman} Coffman V, Kundu  J and Wootters W 2000 {\it Phys. Rev.} A {\bf 61}  052306
\bibitem{osborne06} Osborne T J and Verstraete F 2006 {\it Phys. Rev. Lett.} {\bf 96} 220503
\bibitem{dawson05} Dawson C M, Hines A P, McKenzie R H and Milburn G J 2005 {\it Phys. Rev. A} {\bf 71} 052321
\bibitem{wooters98} Wootters W 1998 {\it Phys. Rev. Lett.} {\bf 80} 2245
\bibitem{melkozernov} Melkozernov A N  and Blankenship R E 2005 {\it Photosynth. Res.} {\bf 85}  33
\bibitem{adolphs06} Adolphs J and Renger T 2006 {\it Biophys. J.} {\bf 91} 2778
\bibitem{wen09} Wen J, Zhang H, Gross M L and Blankenship R E 2009 {\it Proc. Natl. Acad. Sci. USA} {\bf 106} 6134
\bibitem{brixner05} Brixner T, Stenger J, Vaswani H M, Cho M, Blankenship R E and Fleming G R 2005 {\it Nature} {\bf 434} 625
\bibitem{nazir09} Nazir A 2009 {\it Phys. Rev. Lett.} {\bf 103} 146404 
\bibitem{jang08} Jang S, Cheng Y-C, Reichman D R and Eaves J D 2008 {\it J. Chem. Phys.} {\bf 129} 101104 
\bibitem{jang09} Jang S 2009 {\it J. Chem. Phys.} {\bf 131} 164101 



\end{thebibliography}
\end{document}